\documentclass[pra,reprint,letterpaper,superscriptaddress,amsfonts,amsmath,amssymb,nofootinbib,,longbibliography]{revtex4-2}

\usepackage{braket}
\usepackage{graphicx}
\usepackage{color}
\usepackage[normalem]{ulem}
\usepackage{booktabs}
\usepackage{MnSymbol}
\usepackage{enumerate}
\usepackage{bbold}
\usepackage{subfigure}
\usepackage{subfiles}
\usepackage{units}
\usepackage[usenames,dvipsnames]{xcolor}
\usepackage[T1]{fontenc}
\usepackage{caption}
\usepackage{amsmath}
\usepackage{paralist}
\usepackage{psfrag}
\usepackage{setspace,tabularx,fancyhdr}
\usepackage{mathtools}
\usepackage[top=2cm,bottom=2cm,left=2cm,right=2cm]{geometry}
\usepackage{enumitem}
\setlist[itemize]{leftmargin=*}
\usepackage{wrapfig}
\usepackage{type1cm}
\usepackage{yfonts}
\usepackage{relsize}
\usepackage{multirow}
\usepackage{tabularx}
\usepackage{array}
\usepackage{rotating}
\usepackage{bigints}
\PassOptionsToPackage{breaklinks}{hyperref}
\usepackage{xurl} 
\usepackage{hyperref}
\hypersetup{colorlinks=true,linkcolor=MidnightBlue,citecolor=blue,filecolor=green,urlcolor=Violet}

\def\beq{\begin{equation}}
\def\eeq{\end{equation}}
\def\bsp{\begin{split}}
\def\esp{\end{split}}
\def\bea{\begin{eqnarray}}
\def\eea{\end{eqnarray}}

\definecolor{mygreen}{rgb}{0.2, 0.8, 0.2}

\newcommand{\IGNORE}[1]{}

\graphicspath{{Images/}}

\begin{document}

\title{Constraining Dirty Black Holes and pseudo-complex General Relativity \\ with the Gravitational Waves Transient Catalog 3.0}

\author{Yehu I. Maimon}\email{yehu.maimon@biu.ac.il}
\affiliation{Department of Physics, Bar Ilan University, Ramat Gan 5290002, Israel}

\author{Alex B. Nielsen}
\affiliation{Department of Mathematics and Physics, University of Stavanger, NO-4036 Stavanger, Norway}

\author{Ofek Birnholtz}\email{ofek.birnholtz@biu.ac.il}
\affiliation{Department of Physics, Bar Ilan University, Ramat Gan 5290002, Israel}

\begin{abstract}
We use data from the Gravitational Wave Transient Catalog 3.0 to update constraints on parameterized deviations from General Relativity, as encountered in pseudo-complex general relativity (pcGR) theory and models of dirty black holes. The pcGR framework extends Einstein's theory of general relativity by introducing additional parameters that diverge from standard predictions in the strong-field regime, potentially excluding black hole horizons for specific parameter choices. 
We analyze gravitational wave signals from coalescing compact objects to obtain new bounds on these parameters. Our results modify existing constraints and identify previously unexplored regions of parameter space, exploring the observational viability of dirty black holes and horizonless solutions in pcGR.
We confirm the exclusion of 1PN deviations sufficient to avoid a horizon, and for the first time rule out 1.5PN as well.
We also discuss implications for current and future gravitational wave observations in refining these constraints
\end{abstract} 

\maketitle

\section{Introduction}\label{section: Introduction}
General relativity (GR) \cite{EinsteinOG} has been the cornerstone of modern gravitational physics since its inception, providing a remarkably successful description of spacetime and gravity. While GR has been extensively tested in the weak-field regime, its predictions in the strong-field regime — particularly near black hole (BH) horizons — offer fertile ground for exploring potential deviations from standard gravity or standard vacuum solutions. One such approach, is pseudo-complex general relativity (pcGR) \cite{hess2008, hessbook, hess2022,CasperAndHess}, which modifies the geometric framework of GR by introducing additional parameters. 

{pcGR is especially useful for our observational objectives, as it offers a straightforward toy model characterized by two additional parameters, $b$ and $n$.
As elaborated throughout the paper, these parameters provide a means to quantitatively assess how well observations can constrain deviations from GR.
Specifically, $b$ determines the size of the deviation, while $n$ determines the range over which the deviation is significant \cite{Nielsen2017TestingPG,Nielsen:2019ekf}.
We stress that our aim is not to advocate for pcGR as a correct description of nature, but rather to employ it as a convenient framework for placing observational bounds on such deviations.}

While pcGR converges with GR predictions in the weak-field regime  {for large $r$ (or more precisely $(\frac{M}{r})^n$)}, it diverges significantly in the strong-field domain. This divergence allows for solutions that exclude BH horizons, presenting intriguing alternatives to classical BHs.

A different framework, which provides mathematically similar solutions is that of dirty BHs, which represent generic stationary metrics influenced by surrounding matter fields (these are typically not vacuum solutions of the Einstein equations). These models extend beyond vacuum solutions of Einstein’s equations and can accommodate horizonless objects or compact objects with modified spacetime structures.

Observations of gravitational waves (GWs) by the observation run O3b of the Advanced LIGO \cite{adv-ligo}, Virgo \cite{adv-virgo} and KAGRA \cite{kagra,kagra2,kagra3} detectors  {(collectively abbreviated to LVK)}, described in the Gravitational Wave Transient Catalog 3.0 \cite{gwtc3a,gwtc3b}, provide an opportunity to probe the strong-field regime and constrain deviations from GR. Previous studies have demonstrated the potential of gravitational wave signals to place bounds on pcGR parameters, including those related to dirty BH models and horizonless solutions \cite{Nielsen2017TestingPG,Nielsen:2019ekf}.

Despite these advances, significant gaps remain in our understanding of which pcGR parameters, particularly the values governing the theory’s deviations, remain allowed by observable phenomena. This study seeks to address these gaps by leveraging the latest gravitational wave observations to refine and expand constraints on pcGR. By focusing on the parameters, we aim to explore their role in shaping the near-horizon regime and their compatibility with current observational limits.

Our work makes several contributions: (i) it provides updated constraints on pcGR parameters using the most comprehensive gravitational wave dataset to date, (ii) it identifies new regions of parameter space that remain observationally viable, and (iii) it examines the implications of these constraints for the existence of dirty BHs and horizonless solutions.
In Sec \ref{section: BackG}, we provide a brief mathematical background and the main expressions we use for our calculations. In Sec \ref{section:anal}, we present our analysis of selected events, parameters from each of them and calculation of $\delta_{\phi}$. In Sec \ref{section: Results}, we present and explain our results and in Sec \ref{section: Summary}, we summarize our work and discuss future work with future data sets.

\section{Background}\label{section: BackG}
Following \cite{CasperAndHess}, we start with the differential line element $ds^2 = g_{\mu\nu}dx^\mu dx^\nu$  {similarly} to the
Kerr metric \cite{kerr} in Boyer-Lindquist coordinates \cite{boyer-lin} for a BH with mass $M$ and spin parameter $a$, listing all the non-zero metric components:
\beq
\begin{split}
g_{tt} &= -(1-\frac{\psi}{\Sigma}), \quad 
g_{rr} = \frac{\Sigma}{\Delta}, \quad 
g_{\theta\theta} = \Sigma \\
g_{\phi\phi} &= \left( (r^2+a^2)+\frac{a^2\psi}{\Sigma}\sin^2{\theta} \right)\sin^2{\theta} \\
g_{t\phi} &= g_{\phi t} = - a \frac{\psi}{\Sigma}\sin^2{\theta}   ~,
\end{split}
\eeq
where $\Sigma = r^2 + a^2\cos^2\theta$ and $\Delta = r^2 + a^2 - \psi(r)$.
If we set $\psi(r) = 2Mr$, this is the exact GR solution for a Kerr BH solution in vacuum with mass $M$ and spin $a$.
More generally we can write $\psi(r) = 2m(r)r$, where again $m(r)=M$ is the GR Kerr vacuum solution.
A dirty black hole in GR can be described with a different, $r$-dependent mass function $\psi(r)$ (implying a non-vacuum mass-energy distribution); alternatively, the same solution can be interpreted as a possible vacuum solution in pc-GR, where the deviation from GR manifests at leading order $n$ in the ratio $\frac{M}{r}$, i.e. in the form

\bea
m(r) &=& M - \frac{B}{2r^n} = Mg(r), \\
g(r) &=& 1 - b \left(\frac{M}{r}\right)^n ~,
\label{deviation}
\eea
where $b$ is a dimensionless parameter which in vacuum GR equals zero\footnote{The Johannsen-Psaltis (JP) metric \cite{JP,ZaryabAhmedThesis} suggests a somewhat similar non-vacuum modifications of the Kerr metric; expanded in its leading PN order, the JP metric coincides with our ansatz for $a=0$, $\theta=\pi/2$, or both.}.
The parameter $b$ can have a value such that there are no Killing horizons. 
Similarly to \cite{Nielsen2017TestingPG}, we  find the critical $b$ by finding the solutions of

\beq
r^2 + a^2 - 2M\left[1-b\left(\frac{M}{r}\right)^n\right]r = 0 ~.
\eeq

 There are no horizons when the equation does not have real positive solutions, which occurs when $b$ is greater than 

 \beq
 \label{b_c}
 b_{c} = \Upsilon^n \left(1-\frac{\chi^2}{2\Upsilon} - \frac{\Upsilon}{2}\right)    ~,
 \eeq
 where $\Upsilon = (n+\sqrt{n^2-(n^2-1)\chi^2})/(n+1)$  {and $\chi$ is the dimensionless spin parameter}.

\begin{figure}[h]
    \centering
    \includegraphics[width=0.5\textwidth]{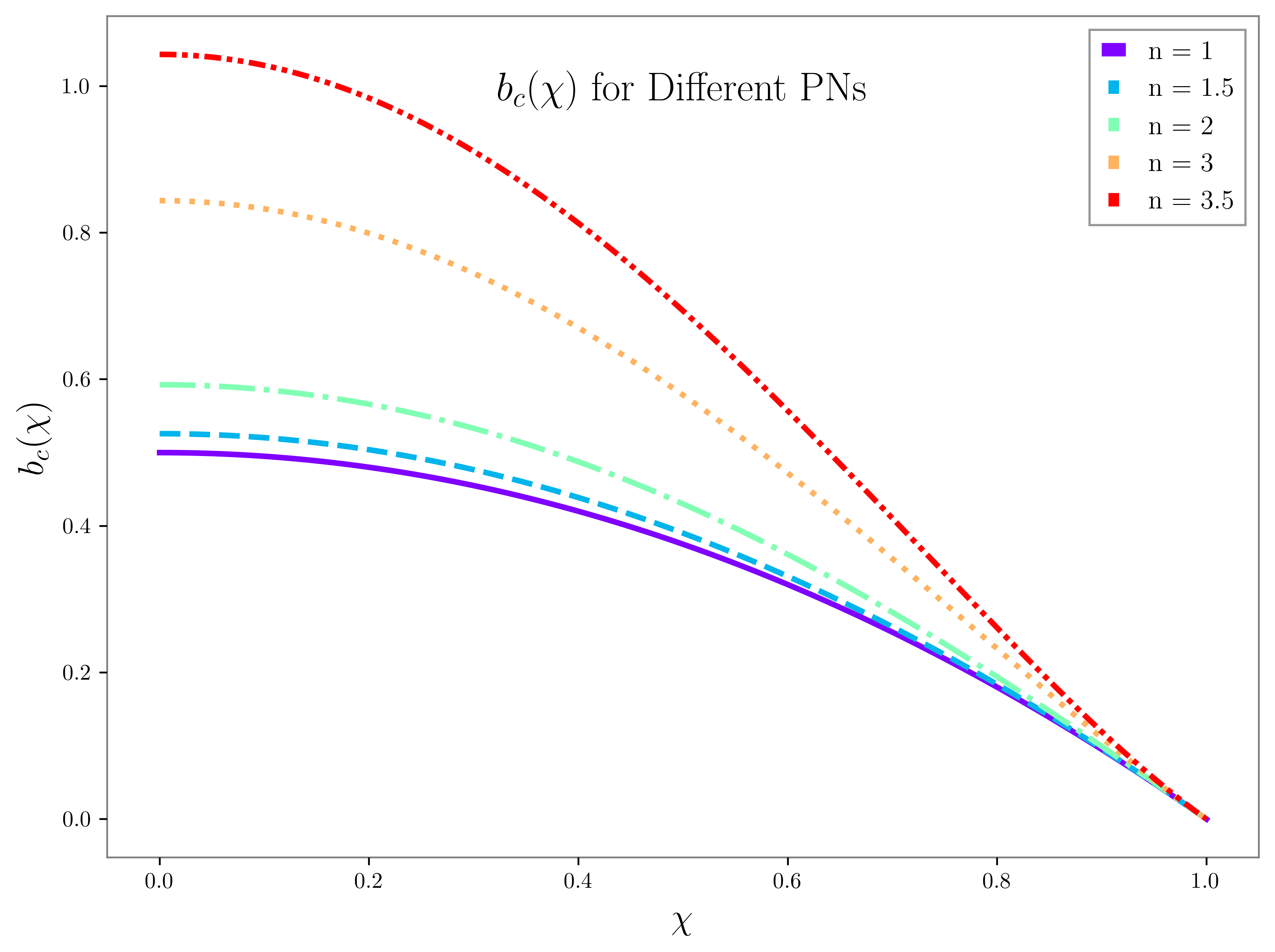}
    \caption{Plot of $b_{c}$ for different values of $\chi$ for different PN orders of deviations in (\ref{deviation}). These curves represent lower bounds for the horizonless case, i.e., at $b>b_c$ no horizon exists.
    In the extremal case, the surface gravity vanishes even at $b=0$.}
    \label{fig:b_anal}
\end{figure}

 Thus, for large enough $b$, black hole horizons do not exist. Also, we can see in Figure \ref{fig:b_anal}, that in the extremal spin case, $b_{c}$ is zero for all $n$'s. We further assume the dimensionless $b_{c}$ is independent of $M$.

 The deviation from GR imprints itself on the gravitational waveform in the following way: as in \cite{Nielsen2017TestingPG,Nielsen:2019ekf} we use the stationary phase approximation method from \cite{cutler}, so that the phase deviation takes the form of 

\begin{equation}
\begin{aligned}
\Psi &= 2\pi f t_c - \phi_c - \frac{\pi}{4} +
\\
&\quad 
\frac{3}{128(\pi G\mathcal{M}f)^{5/3}} \left[1+\sum_{n} \varphi_n +\frac{20}{(n-4)(2n-5)} \mathcal{B}(f) \right]   ~,
\end{aligned}
\end{equation}
where $t_c$ and $\phi_c$ are constants of integration, $\varphi_n$ is the $n$th PN term ($PN_{n}^{GR}$) and \\
\bea
\mathcal{B}(f) 
  &=& \left(\frac{(n+2)(n+1)}{3}Q
     + \rho \right)b(M2\pi f)^{2n/3}, \label{deviation} \\
 Q &=& \frac{1+q^n}{(1+q)^n}~,
\eea
where $q$ is the mass ratio of the progenitor components, $f$ is the frequency of the binary, $\rho$ differentiates between models\footnote{As shown in \cite{Nielsen2017TestingPG,Nielsen:2019ekf}, it favors GW emission by setting it to zero and also it makes it harder to rule out $n$-order in our work where \cite{hessbook} sets it to 1.} and is set to zero for now.

Thus, the following form is comparable with the expected PN coefficients in GR \cite{blanchet,Maggiore,pn_Khan_and_Husa}
\beq
PN_{n}^{pcGR} = \frac{20}{(n-4)(2n-5)} \left(\frac{(n+2)(n+1)}{3}Q + \rho \right)b
\eeq

The next section provides a more in-depth analysis of how we quantify and put constraints on this notion. 

\section{Analysis of GW Events}\label{section:anal}
Following the model presented in
\cite{Nielsen2017TestingPG,Nielsen:2019ekf}, we start by using the O3b data \cite{o3b-data,ligo_scientific_collaboration_and_virgo_2023_8177023}, and analyzing all events from the GWTC-3 catalog \cite{gwtc3b} that met the selection criteria for parameterized PN modification tests \cite{Mehta,Pai:2012mv,PhysRevD.85.082003,PhysRevD.89.082001}, i.e GW191129, GW191204, GW191216, GW200115, GW200129, GW200202, GW200225, GW200311, GW200316\footnote{We use the  shortened event labels, i.e., the date is kept but the 6 digit time is omitted for brevity; no confusion should arise.}
The source parameter inference was conducted using Bilby \cite{bilby_paper}, Parallel Bilby \cite{pbilby_paper,skilling2004,skilling2006,dynesty_paper}, and RIFT \cite{Pankow:2015cra,Lange:2018pyp}, with the results formatted through PESummary \cite{pesummary}.
From these events, we used the averaged posterior  of the final spin $\chi$ of the black hole after the merger, the individual spins of each progenitor $\chi_1$ and $\chi_2$, and the mass ratio $q$. 
We calculate $b_c$ for every event, using \eqref{b_c} with the final spin $\chi$ and the different $n$'s. 
We use the final spin for two reasons: that parameter effectively determines the geometry from the late inspiral through the merger and ringdown, including setting the innermost stable circular orbit (ISCO), the light ring (LR), and the final horizon itself \cite{Basic_Physics, McWilliams:2018ztb, Buskirk:2018ebn}; in the early inspiral, when the individual constituent spins are expected to be more important, we notice that for most LVK events (and all events we consider) the final spin is larger than any of the individual ones, and thus poses an even more restrictive bar on distinguishing $b_c$.

\begin{figure}[h]
    \centering
    \includegraphics[width=0.85
    \textwidth]{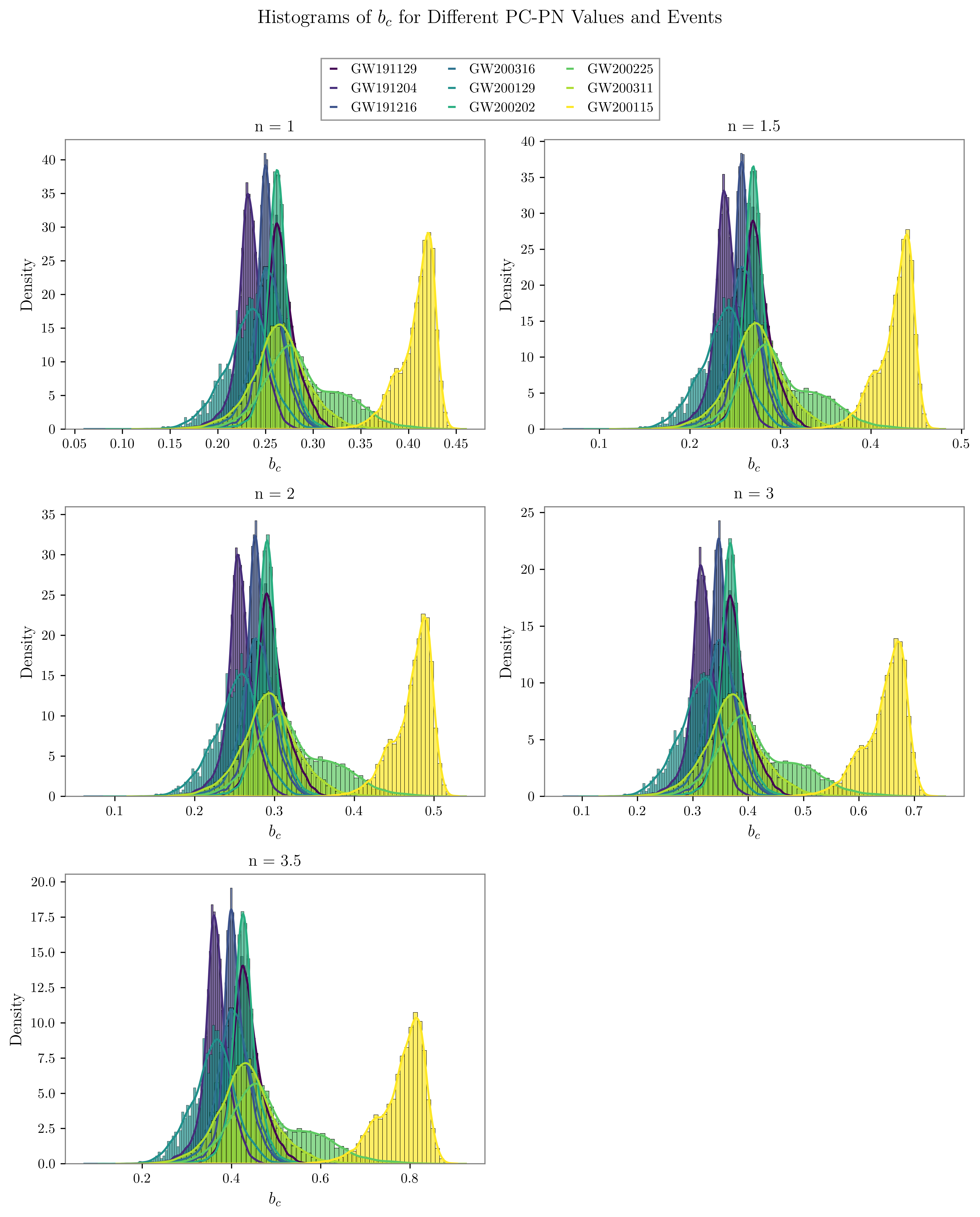}
    \begin{minipage}{\textwidth}
        \captionsetup{justification=centering, singlelinecheck=false}
        \caption{Histograms of the posterior densities of $b_c$ for different $PN^{pc}$ values from all 9 events. As shown in Fig. \ref{fig:b_anal}, $b_c$ depends strongly on the final spin, and thus the events' $b_c$'s are seen to group by final spin values. All events except GW200115 are consistent with a spin of 0.7; GW200115, with a lower spin, projects a higher $b_c$.  }
        \label{fig:b_crit}
    \end{minipage}
\end{figure}

Figure \ref{fig:b_crit} presents histograms of $b_{c}$ for various values of $n$. Notably, the event GW200115 exhibits a different histogram compared to the others and was treated separately.
This difference can be attributed to its averaged final spin, which is $0.426$, whereas the remaining events all have final spin values around $0.7$.
Then we calculate the post-Newtonian (PN) terms, the calculations are shown in our code on GitHub \cite{ymgit}, of pseudo-complex general relativity (pcGR) and general relativity (GR) as mentioned in Sec \ref{section: BackG}, and compute the ratios $\delta_\varphi$ as:
\beq
\delta_{\varphi} = \frac{PN_n^{pcGR}}{PN_n^{GR}}
\eeq
Using these ratios, we determine whether $PN_n^{pcGR}$ is significant compared to $PN_n^{GR}$ and letting us determine which $n$'s can be ruled out.
The boundaries for $PN_n^{GR}$ represent observational constraints, which means that $PN_n^{pcGR}$ must exceed these boundaries to be ruled out. Applying the GWTC-3 limits, we can rule out any values of $PN_n^{pcGR}$ that exceed the limits of $PN_n^{GR}$. The calculations are shown in Table \ref{table:pn_val}.

Our code is openly available at \cite{ymgit}.

\section{Results}\label{section: Results}

\begin{figure}[h]
    \centering
    \includegraphics[width=0.85\textwidth]{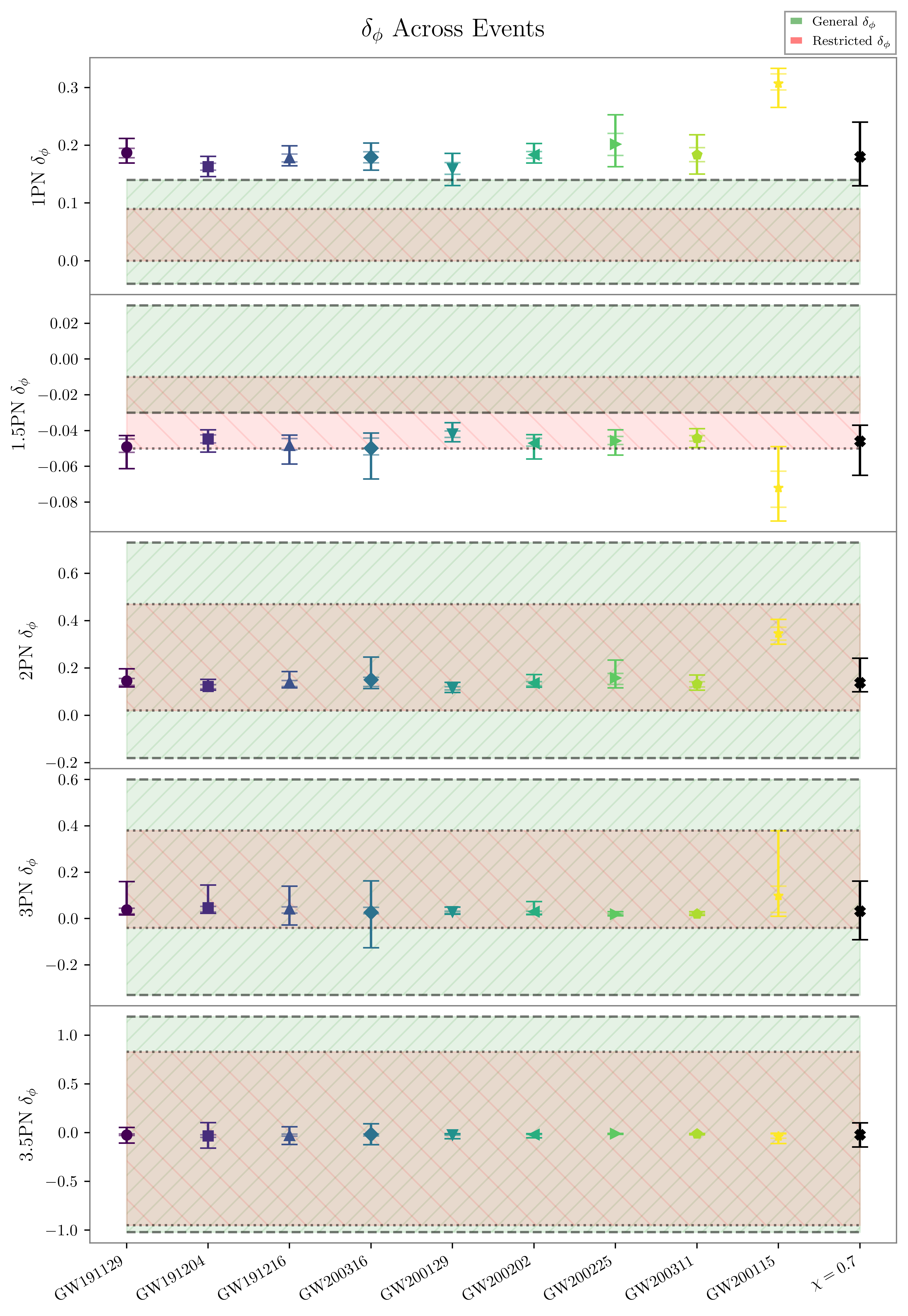}
    \begin{minipage}{\textwidth}
        \captionsetup{justification=centering, singlelinecheck=false}
        \caption{Post-Newtonian (PN) Parameter Plot for GW Events, showing both the $50\%$ (solid bars) and $90\%$ (transparent bars) credible intervals across the events. We also show the values of the averaged $\chi$ according to Table \ref{table:pn_val}. 
        The constraints found by general variations of the PN parameters (independently for different events) are marked as a lightgreen shaded band with bottom-left to top-right green diagonal lines; the constraints found by restricting the deviations to be the same across all events are marked as a pink shaded band with top-left to botton-right red diagonal lines; the restricted band mostly overlaps with (or is included within) the general band.}
    \label{fig:pn_plot}
    \end{minipage}

\end{figure}

Following Section \ref{section:anal}, we examine the error limits of GWTC-3 for each PN order (1, 1.5, 2, 3, and 3.5). Figure \ref{fig:pn_plot} presents the results across events for all PN terms and also the $\chi = 0.7$ case is added. We highlight the General (green) and Restricted (red) constraints on the deviation parameter $\delta_{\varphi}$. The General (Restricted) constraints assume that deviation coefficients can (cannot) vary across observed events.

 {While the LIGO collaboration tests post-Newtonian (PN) coefficients by directly comparing observational data to the predictions of general relativity (GR), searching for deviations from the theory, our approach differs slightly. We compute the ratio of the PN coefficients predicted by pcGR and GR, and then compare the observationally inferred values to this theoretical ratio. This allows us to quantify how much pcGR deviates from GR relative to the observations, rather than testing each theory in isolation.}

We see that the 1PN and 1.5PN deviations are strongly excluded.
The 1PN deviation lies outside both the General and Restricted bounds, while the 1.5PN deviation - tested here for the first time - lies entirely outside the General bounds.
The other PN deviations, 2PN, 3PN and 3.5PN are all within the boundaries and thus are not ruled out.
The event GW200115 has a noticeably different final spin, so we treated it as a single event rather than part of a series of similar events when analyzing the data. It provides the most stringent exclusion bound for 1PN and 1.5PN, and is the closest to the exclusion boundary for 2PN as well.\\

\section{Summary}\label{section: Summary}
In this paper, we have explored and calculated the pseudo-complex Post-Newtonian (pcPN) terms and compared them to the observed PN constraints from the GWTC-3 catalog \cite{gwtc3a}.
Our analysis indicates that the 1PN and 1.5PN orders can be considered effectively ruled out based on our calculations and the current data.
We used a total of 9 events of O3b, from the GWTC-3 catalog -- compared to the previous catalog, GWTC-2 \cite{gwtc2}, which contains 25 events used for parameterized tests, albeit providing altogether less restrictive constraints as detector sensitivities, and the associated typical signal-to-noise ratios (SNRs), were lower.
The combined bounds used in \cite{Nielsen2017TestingPG,Nielsen:2019ekf}, were based on the events GW150914 and GW151226 \cite{gw150914} and these bounds for 1PN, 2PN and 3PN were $(-20\%,5\%)$, $(-130\%,15\%)$ and $(-100\%,10\%)$\footnote{The 3PN bound in \cite{Nielsen2017TestingPG,Nielsen:2019ekf} was erroneously conflated with the 3PNlog bounds $(-100\%,600\%)$; $(-100\%,10\%)$ is the correct range from the O1 events} respectively.
The new bounds we have presented are narrower in their constraints - albeit, for the 2PN and 3PN deviations, they have shifted in a direction more accommodating of pcGR horizonless objects.
We also note that the model used in \cite{Nielsen2017TestingPG,Nielsen:2019ekf} assumed a mass ratio $q=1$ and no spin, an assumption which increased the pcGR deviations expected relative to the deviations appropriate for the events we use from GWTC-3 - mostly because as the final spin approaches 1, a smaller deviation is required to avoid the horizon (as in Fig. \ref{fig:b_anal}).

In future work, as suggested in \cite{hess_ray,hess_lr1,hess_lr2}, we can explore and test more bounds on pcGR using different observation regions for modified light-ring structures which can be tested using The Event Horizon (EHT) survey results  \cite{Psaltis_2019} and from the GW perspective we can use the equivalent of testing the light-ring using Quasi-Normal Modes (QNMs) \cite{Berti_2006,Yang_2012}.  {Our analysis of QNMs will begin with a revision of the eikonal limit in pcGR, followed by a comparison with the corresponding results in GR and with observational data, as was done by \cite{Isi_2019} for GW150914 (thus far finding agreement). It is also worth noting that horizonless compact objects can still possess a photon ring, as discussed in \cite{Nielsen2017TestingPG}, so we will examine how far the parameters $n$ and $b$ can be pushed without breaking the QNM-LR agreement.}
 {We plan to calculate how various QNM modes in pcGR differ for different values of $n$, starting with the fundamental mode and the first overtone, primarily considering the Eikonal limit. This work will differ from the current study, which focuses on the inspiral phase; in future work, we will focus on the ringdown phase. Our overarching goal is to constrain the pcGR parameters $b$ and $n$.}

After utilizing the GWTC-3 catalog for our analysis, we plan to update our results upon the release of catalogs of events from the O4 observation runs.
If detector sensitivities and SNRs improve \cite{LIGO_G2002127}
this should allow us to refine our constraints and perform a renewed test of pseudo-complex general relativity (pcGR) and dirty black holes using the latest data.

\begin{widetext}
\begin{table*}
\begin{center}
\begin{tabular}{c c c c c c c c}
    \toprule
        $n$ & $\chi$ & $b_{\text{crit}}$ & $\text{PN}^{\text{(pc-GR)}}$ & $\text{PN}^{\text{(GR)}}$ & $\delta_{\varphi}$ & General & Restricted \\
        \midrule
        1  & 0 & 0.5  & 2.22 & 6.44 & 34$\%$ & \multirow{4}{*}{(-4$\%$, 14$\%$)} & \multirow{4}{*}{(0$\%$, 9$\%$)} \\
          & 0.426 &  $0.41^{0.43}_{0.37}$ & $1.8^{1.91}_{1.65}$ & $5.92^{6.38}_{5.6}$ & $30.7\%^{33\%}_{26.6\%}$  & & \\
          & 0.7 & $0.257^{0.34}_{0.19}$  & $1.14^{1.52}_{0.86}$ & $6.35^{6.44}_{5.91}$ & $18\%^{24\%}_{13\%}$ & & \\  
          \hline
        1.5 & 0  & 0.526  & 4.33 & -50.26 & -8.6$\%$ & \multirow{4}{*}{(-3$\%$, 3$\%$)} & \multirow{4}{*}{(-5$\%$, -1$\%$)} \\
          & 0.426 & $0.426^{0.45}_{0.38}$  & $3.98^{4.38}_{3.3}$ & $-56.36^{-47.5}_{-68.2}$ & $-7.2\%^{-4.9\%}_{-9.1\%}$  & & \\
          & 0.7 & $0.264^{0.35}_{0.2}$  & $2.23^{3.02}_{1.67}$ & $-48.22^{-38.76}_{-60.76}$ & $-4.6\%^{-3.7\%}_{-6.5\%}$ & & \\
        \hline
        2  & 0  & 0.592  & 11.84 & 46.24 & 25.61$\%$ & \multirow{4}{*}{(-18$\%$, 73$\%$)} & \multirow{4}{*}{(2$\%$, 47$\%$)} \\
          & 0.426 &  $0.47^{0.5}_{0.43}$ & $12.68^{14.8}_{9.1}$ & $36.5^{38.6}_{29.6}$ & $34.6\%^{40.5\%}_{30\%}$  & & \\
          & 0.7 & $0.284^{0.39}_{0.21}$  & $6.01^{8.6}_{4.42}$ & $44.14^{46.2}_{33.6}$ & $13.8\%^{24.2\%}_{10\%}$ & & \\
        \hline
        3  & 0  & 0.84  & -28.1 & -652 & 4.31$\%$ & \multirow{4}{*}{(-33$\%$, 60$\%$)} & \multirow{4}{*}{(-4$\%$, 38$\%$)} \\
          & 0.426 & $0.646^{0.69}_{0.57}$  & $-43.6^{-22.3}_{-57.5}$ & $-755.8^{-28.2}_{-1745.4}$ & $9.9\%^{37.9\%}_{1.1\%}$  & & \\
          & 0.7 & $0.359^{0.51}_{0.25}$  & $-14.01^{-9.3}_{-25.2}$  & $-474.2^{-355.6}_{-1278.5}$ & $3.2\%^{16.2\%}_{-9.1\%}$ & & \\
        \hline
        3.5  & 0  & 1.04  & -30.4 & 1129.2 & -2.7$\%$ & \multirow{4}{*}{(-102$\%$, 119$\%$)} & \multirow{4}{*}{(-95$\%$, 83$\%$)} \\
          & 0.426 &  $0.782^{0.84}_{0.69}$ & $-57.8^{-24.7}_{-80.1}$ & $2108.4^{3961.3}_{501.9}$ & $-3.7\%^{-0.6\%}_{-11.2\%}$  & & \\
          & 0.7 & $0.415^{0.61}_{0.29}$  & $-15.16^{-9.3}_{-31.36}$ & $790^{2800.1}_{-1060.3}$ & $-2.1\%^{10\%}_{-14.6\%}$ & & \\
        \hline
    \bottomrule
\end{tabular}
\caption{Parameter calculations using GWTC-3, divided into three cases for each PN order. The parameters $q$ and $\chi$ were taken from the GWTC-3 events and used in our analysis.
For $\chi = 0$, the mass ratio $q$ is set to 1, and the results are compared with those from \cite{Nielsen2017TestingPG}. For $\chi = 0.426$, the event GW200115, which has a $\chi$ value that is distinct from the rest of the events and considered as an outlier, is used. For $\chi = 0.7$, the remaining events are considered. Each calculated parameter has two boundaries, which correspond to the 90\% credible interval,
and is given as $\texttt{median}^{\texttt{upper 90\%}}_{\texttt{lower 90\%}}$
. The General (Restricted) constraints on $\delta_{\phi}$ assume that deviation coefficients can (cannot) vary across observed events.}
\label{table:pn_val}
\end{center}
\end{table*}
\end{widetext}


\clearpage

\clearpage

\section*{Acknowledgments}\label{section: Acknowledgments}
We acknowledge support from the US-Israel Binational Science Fund (BSF) grant No. 2020245 and the Israel Science Fund (ISF) grant No. 1698/22. We also thank the annonymous referee for useful comments. \\
In addition, This research has made use of data or software obtained from the Gravitational Wave Open Science Center (gwosc.org), a service of the LIGO Scientific Collaboration, the Virgo Collaboration, and KAGRA. This material is based upon work supported by NSF's LIGO Laboratory which is a major facility fully funded by the National Science Foundation, as well as the Science and Technology Facilities Council (STFC) of the United Kingdom, the Max-Planck-Society (MPS), and the State of Niedersachsen/Germany for support of the construction of Advanced LIGO and construction and operation of the GEO600 detector. Additional support for Advanced LIGO was provided by the Australian Research Council. Virgo is funded, through the European Gravitational Observatory (EGO), by the French Centre National de Recherche Scientifique (CNRS), the Italian Istituto Nazionale di Fisica Nucleare (INFN) and the Dutch Nikhef, with contributions by institutions from Belgium, Germany, Greece, Hungary, Ireland, Japan, Monaco, Poland, Portugal, Spain. KAGRA is supported by Ministry of Education, Culture, Sports, Science and Technology (MEXT), Japan Society for the Promotion of Science (JSPS) in Japan; National Research Foundation (NRF) and Ministry of Science and ICT (MSIT) in Korea; Academia Sinica (AS) and National Science and Technology Council (NSTC) in Taiwan.
Some of the results in this paper have been derived using the ``PESummary``
package.
This research has made use of Parallel Bilby, a parallelised Bayesian inference Python package, and Dynesty, a nested sampler, to perform Bayesian parameter estimation and RIFT, Rapid parameter inference on gravitational wave sources.



%

\end{document}